\documentclass[aps,prl,twocolumn,groupedaddress,showpacs]{revtex4-1}
\usepackage{graphicx}  
\usepackage{dcolumn}   
\usepackage{bm}        
\usepackage{amssymb,amsfonts,amsmath}   
\usepackage{color}
\usepackage{lipsum}
\usepackage{upgreek}

\usepackage[normalem]{ulem}

\renewcommand{\d}{\mathrm{d}}

\newcommand{\bs}{\boldsymbol}
\renewcommand{\d}{\mathrm{d}}

\newcommand{\x}{{\boldsymbol x}}
\renewcommand{\r}{{\boldsymbol r}}
\newcommand{\y}{{\boldsymbol y}}
\newcommand{\n}{{\boldsymbol n}}
\newcommand{\f}{{\boldsymbol f}}
\newcommand{\U}{{\boldsymbol U}}
\newcommand{\G}{{\boldsymbol G}}
\renewcommand{\O}{{\boldsymbol \Omega}}

\renewcommand{\u}{{\boldsymbol u}}
\allowdisplaybreaks

\begin{document}

\title{Filter-feeding, near-field flows, and the morphologies of colonial choanoflagellates}

\author{Julius B. Kirkegaard and Raymond E. Goldstein}
\affiliation{Department of Applied Mathematics and Theoretical Physics, Centre for Mathematical Sciences, \\ University of 
Cambridge, Wilberforce Road, Cambridge CB3 0WA, United Kingdom}

\date{\today}


\begin{abstract}
Efficient uptake of nutrients from the environment is an important component in the fitness of all microorganisms, 
and its dependence on size may reveal clues to the origins of evolutionary transitions to multicellularity.  Because
potential benefits in uptake rates must be viewed in the context of other costs and benefits of size, such as
varying predation rates and the increased metabolic costs associated with larger
and more complex body plans, the uptake rate itself is not necessarily that which is optimized by evolution.  
Uptake rates can be strongly dependent on local organism geometry and its swimming speed, providing selective pressure for
particular arrangements.  Here we examine these issues for choanoflagellates, 
filter-feeding microorganisms that are the closest relatives of the animals.
We explore the different morphological variations of the choanoflagellete \textit{Salpingoeca rosetta},
which can exist as a swimming cell, a sessile thecate cell, and as colonies of cells in various shapes.
In the absence of other requirements and in a homogeneously nutritious environment, we find that
the optimal strategy to maximize filter feeding by the collar of microvilli is to swim fast, which favours swimming unicells.
In large external flows, the sessile thecate cell becomes advantageous.
Effects of prey diffusion are discussed and also found to be to the advantage of the swimming unicell.
\end{abstract}

\maketitle

\section{Introduction}
\begin{figure}[b]
\centering
\includegraphics[width=0.48\textwidth]{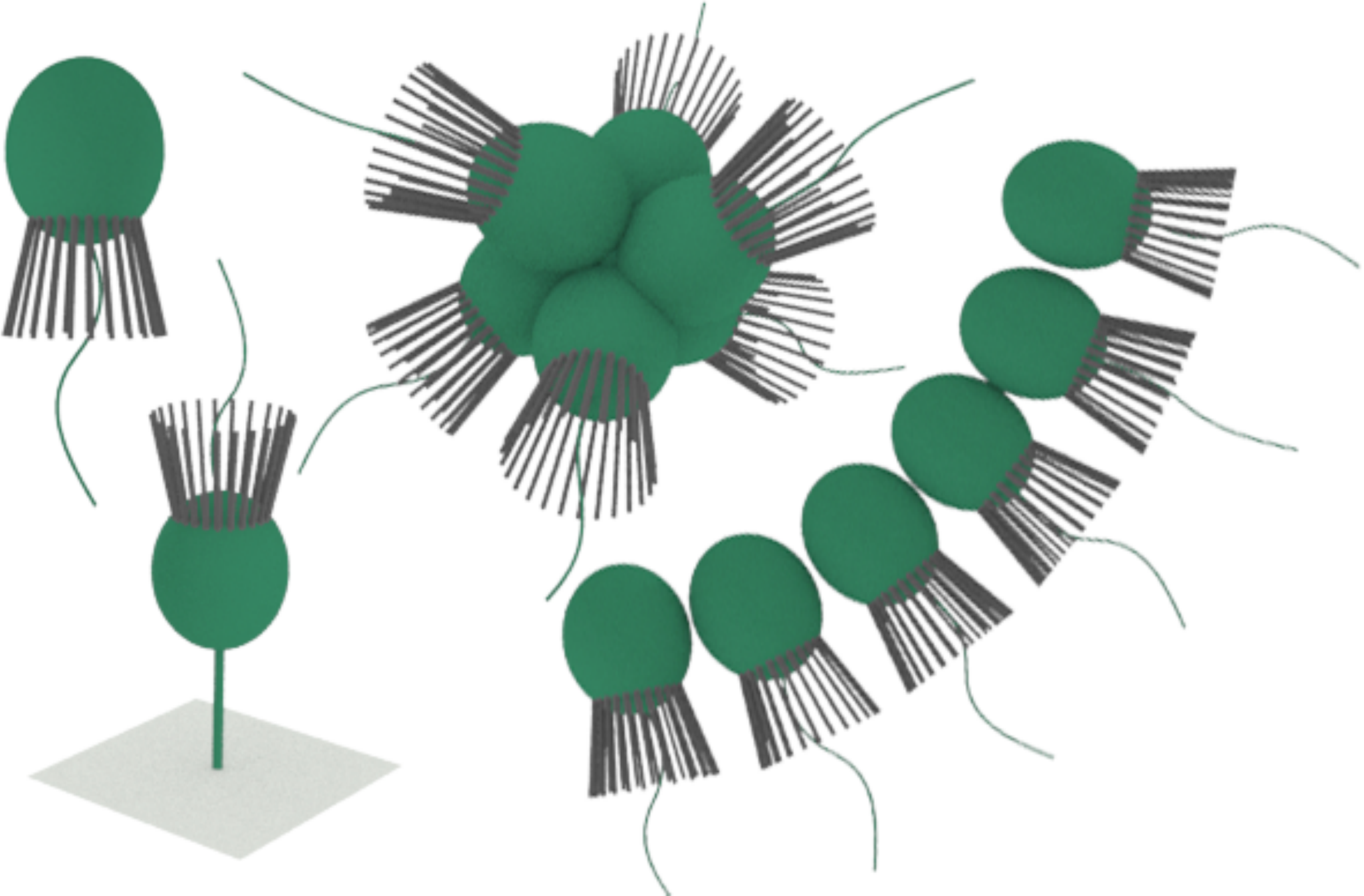}
\caption{Morphologies of \textit{S. rosetta} \cite{Dayel2011} considered here.  
From left to right: swimming unicell, thecate cell, rosette colony and chain colony.}
\label{fig:choano}
\end{figure}

Competitive advantages over single cells is one of the driving forces behind the existence of multicellular life forms.
Certain single-celled organisms mimic true multicellular behavior by forming colonies.
While such colonies do not have the advantages that accrue with division of labor, they do obtain potential benefits
from their increased size, otherwise limited by intracellular nutrient
mixing by diffusion.
In the closest relatives of animals, the choanoflagellates, the species \textit{Salpingoeca rosetta}
can form colonies of both chain-like and rosette-like morphologies \cite{Dayel2011, Alegado2012} as illustrated in Fig. \ref{fig:choano}.
Given their position relative to the origins of animal multicellularity, the possible competitive advantage of these colonies is highly intriguing \cite{Roper2013}.

Choanoflagellates \textit{filter feed} by beating their flagella and thereby driving fluid through a collar of 
microvilli onto which prey (bacteria) get trapped and ingested.
They live at low Reynolds numbers \cite{Purcell2006}, are when swimming freely are 
force- and torque-free, and the surrounding flow $\u$ obeys the Stokes equations
\begin{equation}
\mu \nabla^2 \u = \nabla p, \, \, \nabla \cdot \u = 0,
\end{equation}
where $p$ is the pressure field and $\mu$ the dynamic viscosity.
Being neutrally buoyant, the far-field flow around both unicells and colonies of choanoflagellates is dominated by 
the stresslet contribution which decays as $r^{-2}$ \cite{Kim2005}.
The advective influx of fluid through a sphere of radius $r$ is thus independent of $r$ as $r \rightarrow \infty$.
Using this result, recent work \cite{Roper2013} showed that certain morphologies of colonies such as chains 
can increase this flux per constituent cell,
thus potentially creating a \textit{hydrodynamic feeding advantage} for colonies, in a parallel to the
situation previously examined for the green alga {\it Volvox} \cite{Short2006,Solari2013}.
For choanoflagellates with $n$ constitutent cells, the influx $f$ was shown to grow faster than linearly with $n$ already from $n=2$ and even in the limit $n \rightarrow \infty$ \cite{Roper2013}.

\begin{figure*}
\centering
\includegraphics{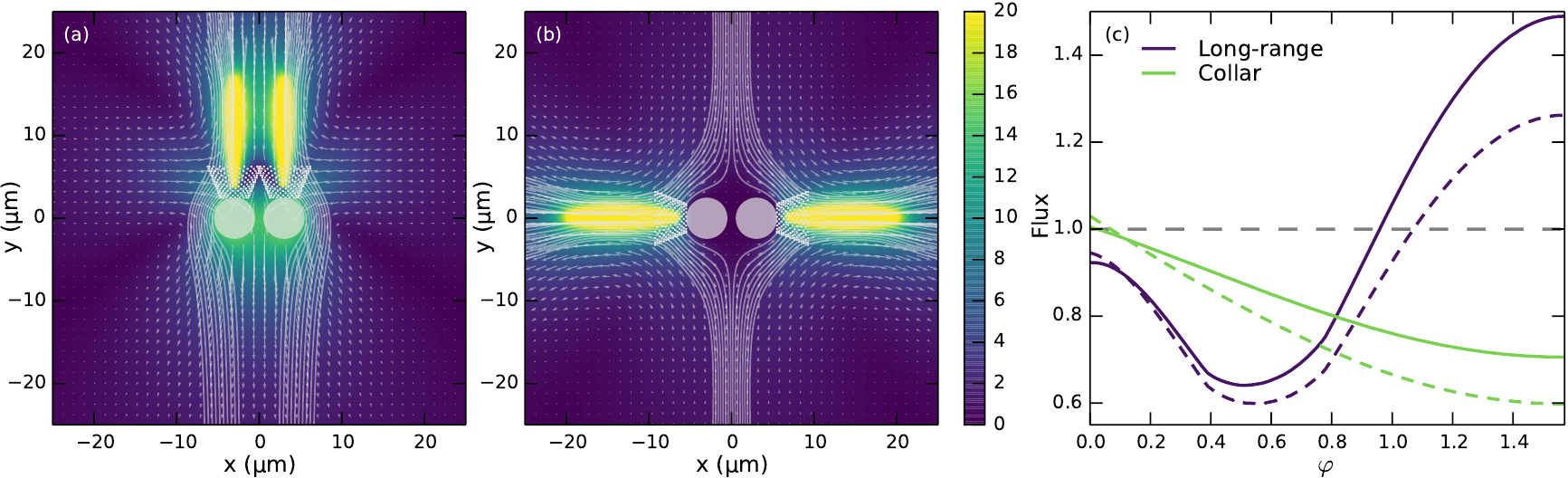}
\caption{Fluid flow and flux of dimers. (a-b) Background color and vector field quantify the velocity field 
in the laboratory frame, with color scale in units of $\upmu$m/s. Streamlines are calculated in the swimming frame 
with $z=0$. Configuration (a) has $\varphi = 0$ and (b) $\varphi = \pi/2$. (c) Influx through a sphere of radius $R \rightarrow \infty$ (neglecting advective flux) shown in purple and flux through the collar of cells in green. Fluxes are calculated per cell and normalised by the flux of a unicell. Solid lines are calculated with velocity described 
on the flagella and dashed lines with forces prescribed on the flagella.}
\label{fig:dimerflow}
\end{figure*}

Theoretically, filter feeding is possible even in the absence of diffusion of the target particles.  In the contrasting case of
absorbers, feeding occurs across a thin diffusive boundary layer, as has
been studied in squirmer-type models \cite{Magar2005,Short2006} consisting of spheres with imposed tangential
velocity fields.
For squirmers it has been shown that optimal nutrient uptake precisely corresponds to optimal swimming, at all P{\'e}clet numbers \cite{Lauga2011a}.
If this result carries over to the filter feeding of colonies of choanoflagellates,
it would suggest that optimally swimming colonies would also be optimal feeding, in constrast to conclusions made based on long-range fluxes \cite{Roper2013}.
Inspired by these studies, we examine here theoretically the near-field flows around colonial choanoflagellates
and the near-field fluxes across the cell collars where feeding occurs. 

\section{Model}
We approximate the choanoflagellate body by a sphere.
Contrary to the prokaryotic case, eukaryotic flagella,
such as those of choanoflagellates, are not rigid rotating helices but instead
their shape is modulated by distributed molecular motors along the flagella to yield a whip-like beat.
Time-averaging over this beat yields an approximately straight line, which we will take as a proxy for the flagella.
Although there is evidence of some influence on the flow of the collar, via a pressure drop across it \cite{Pettitt2002}, we will ignore the collar in the modelling of the fluid flow.
Our system of (colonies of) choanoflagellates thus consists only of spheres and straight lines.

To calculate the flow in unbounded domains, 
we utilizing a boundary element method.
Cortez el. al. \cite{Cortez2005} found the Stokes flow due to a regularized, localized forcing
\begin{equation}
\mu \nabla^2 \u - \nabla p = \delta^\epsilon(r) \, \f = \frac{15 \epsilon^4}{8 \pi (r^2 + \epsilon^2)^{7/2}} \, \f,
\end{equation}
where $r = |\x - \x_0|$ and $\delta^\epsilon$ is a regularized version of the Dirac delta function.
The solution,
\begin{align} \nonumber
\u(\x) &= \frac{(r^2 + 2 \epsilon^2) \, \f + \f \cdot (\x-\x_0) \, (\x-\x_0)}{8\pi \mu \, (r^2 + \epsilon^2)^{3/2}}\\ 
&\equiv \G^\epsilon(\x-\x_0) \cdot \f,
\label{eq:regstokeslet}
\end{align}
is called the regularized Stokeslet, and indeed tends to the classic, singular Stokeslet as $\epsilon \rightarrow 0$.
The flow around a set of boundaries $D$ in an infinite domain can then be approximated by the boundary integral equation \cite{Cortez2005}
\begin{equation}
\u(\x) = \iint_D \G^\epsilon(\x-\x') \cdot \f(\x') \, \d S
\label{eq:bem1}
\end{equation}
with a suitable choice of $\epsilon(\x)$.

Inspired by spectral methods, and as detailed in the appendix, we expand the force distribution on the flagellar elements in terms of Legendre polynomials and on cell bodies in terms of spherical harmonics.
Boundary conditions are no-slip on the cell bodies.
For the flagella boundary conditions we consider two cases.
In both of these cases we take a constant velocity along the flagella: $\u = u_0 \, \hat{\bs d}_i$ (but other velocity distributions could easily be applied).
$u_0$ may then be regarded as known or we can let $u_0$ be indirectly defined by letting the total propulsive force $\f_0 \cdot \hat{\bs d}$ that the flagellum exerts on the fluid be known.
These two choices lead to similar behavior for single cells, but will matter in the case of colonies.
The method detailed in the appendix yields the surrounding flow $\u$, and the translational and rotational swimming velocities, $\U$ and $\O$.

\begin{figure*}
\centering
\includegraphics{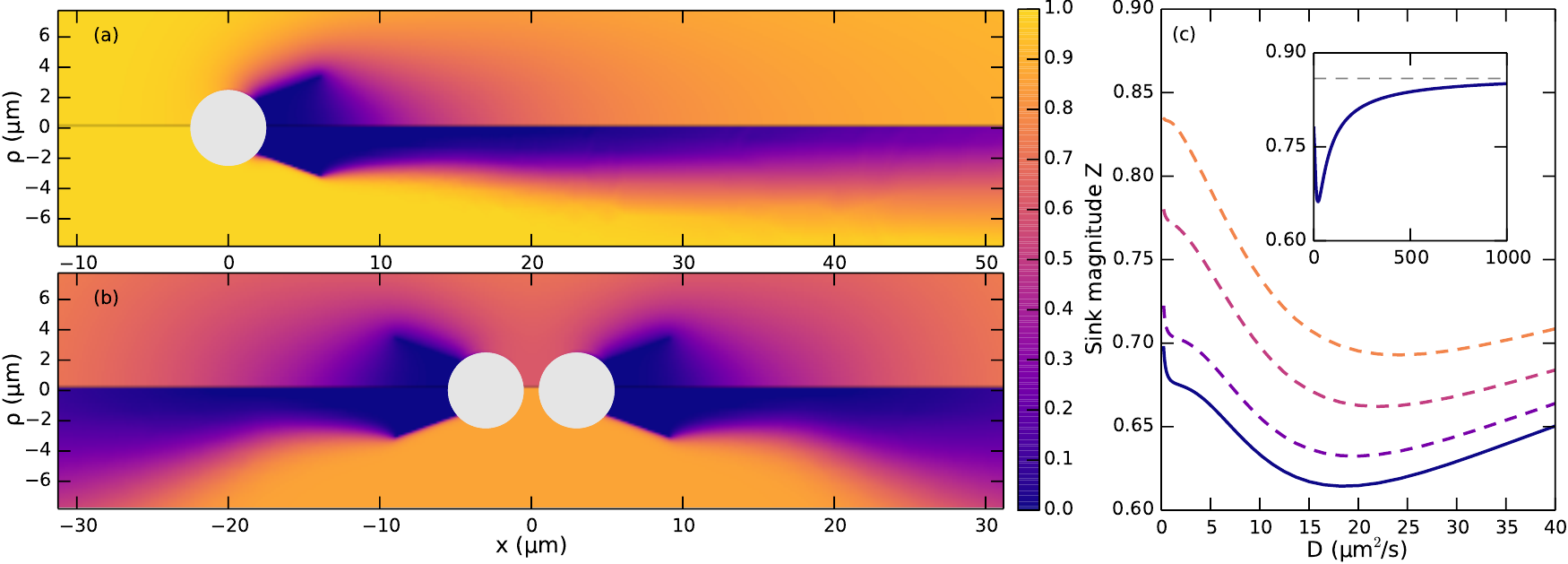}
\caption{Nutrient field. (a) Axially symmetric field around swimming unicell at $D = 1 \, \upmu \text{m}^2/\text{s}$
(lower part) and $D = 25 \, \upmu \text{m}^2/\text{s}$ (upper part). Field is normalised to $c_0 = 1$ at $r \approx 10$
mm. (b) Field around pole-to-pole dimer colony for the same values of $D$ with velocity prescribed boundary condition 
on the flagella. (c) Ratio of $Z$ for a cell in the dimer configuration to swimming unicell. $D=0$ corresponds to 
infinite P{\'e}clet number and compares to the flux calculations shown in Fig. \ref{fig:dimerflow}. Solid curve corresponds 
to configurations shown in (a,b) with collar opening angle $\sim 20^\circ$. Dashed lines for lower collar opening angles, 
with the top (orange) curve corresponding to straight collars. Inset shows a larger range of $D$ and the 
asymptote of vanishing P{\'e}clet number (dashed).}
\label{fig:nutrientfield}
\end{figure*}

\section{Flow around Dimers}

We begin by considering dimers: colonies consisting of two cells.
The two can be placed in various relative orientations; we
focus here on the subset of configurations in which the flagella lie in a plane and where both flagella make the same angle $\varphi$ with the $y$-axis, 
since these are optimal configurations under variation of the remaining angles.
Figures \ref{fig:dimerflow}a,b shows the resulting flow fields for $\varphi=0$ and $\varphi=\pi/2$, respectively.
For $\varphi = 0$ the colony is swimming and the streamlines of passive tracers pass from the front of the colony 
to the back, while for $\varphi=\pi/2$ the forces of the two beating flagella exactly cancel and the colony does not swim.
Passive tracers are dragged in from the sides. For all $\varphi$, $\O = {\bs 0}$.
From these calculations we can reproduce qualitatively the results of Ref. \cite{Roper2013}.
The long-range flux produced by colonies is given by
\begin{equation}
f = \lim_{R \rightarrow \infty} \iint_{S_R, \u \cdot \hat{\n} > 0} \u(\x) \cdot \hat{\n}(\x) \, \d S,
\end{equation}
where $S_R$ is the surface of a sphere with radius $R$ and $\hat{\n}$ is the inward normal to this surface.
The flux per cell, normalised by the flux for the single cell system, is shown in purple in Fig. \ref{fig:dimerflow}c.
Solid curves is the case where $u_0$ is prescribed and dashed is the case $\f_0 \cdot \hat{\bs d}$ prescribed.
Both cases have $\varphi = 0$ and $\varphi = \pi/2$ as local optima, the latter being globally optimal.
This long-range flux per cell is larger than that of a single cell in the pole-pole configuration as previously found \cite{Roper2013}, 
although we find an overall lower magnitude of this long-range flux, due to hydrodynamic interactions between the two cells and differences in geometry choices (\textit{e.g.} distance between cells).
We furthermore find an increased flux in the case of prescribed velocity over prescribed force.

The near-field flow enables us also to calculate the flux not just of an infinite sphere, but also at the actual collars where the choanoflagellates feed.
Evaluating such fluxes allows for the inclusion of the flux due to swimming at speed $\U$.
While earlier work \cite{Roper2013} found that this advective flux was negligible,
this conclusion was based on use of the stresslet flow, which is only valid far away from the colony.
Secondly, although the advective flux may be small compared to the rest of the flux in a particular system,
it can still be important when evaluating the relative flux between systems, as indeed turns out to be the case here.
We find that including the advective flux makes a significant change, and as shown in green in Fig. \ref{fig:dimerflow}c by including this in the flux calculation across the collar,
the swimming side-by-side configuration becomes globally optimal, and the advantage over single cells of colonies disappears (but does not become disadvantageous in the optimal configurations).
Moreover, this behavior is not strongly dependent on the shape of the collar or the distance between the two cells,
in sharp contrast to the long-range flux, the value of which tends to infinity as this distance is increased.
We also find that the prescribed force side-to-side system outperforms single cells slightly due to drag cancellation.

\begin{figure*}
\centering
\includegraphics{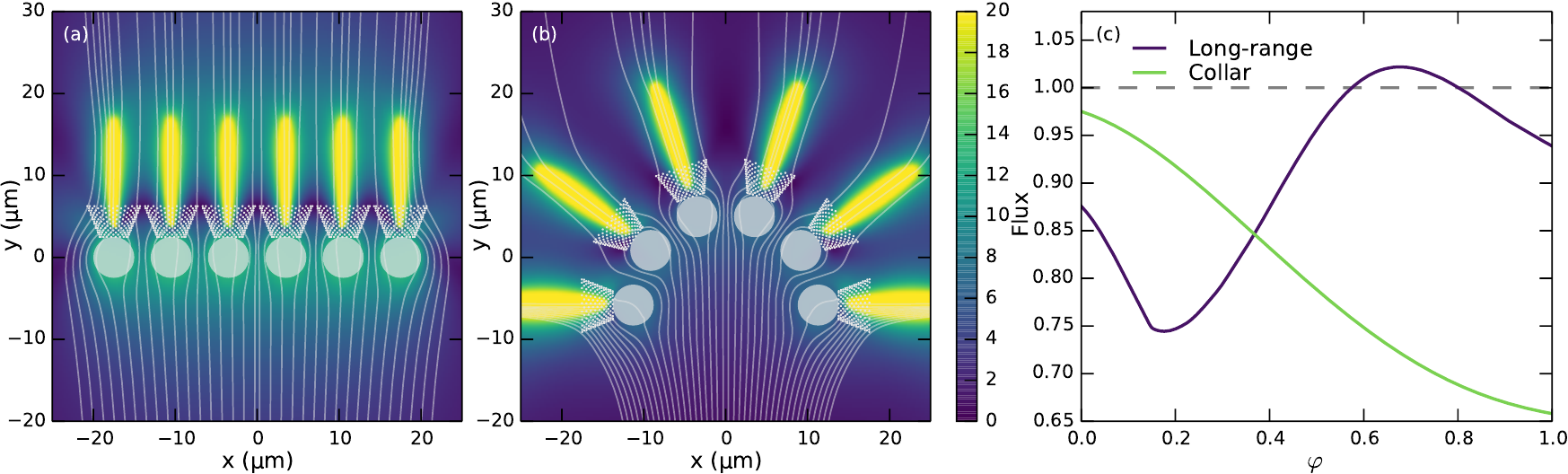}
\caption{Fluid flow and flux of chain colonies. (a-b) Background color shows the magnitude of the velocity field in the 
laboratory frame, with color scale in units of $\upmu$m/s. Streamlines are calculated in the swimming frame from 
$y=-20 \, \upmu$m, $z = 0.1 \, \upmu$m and projected onto the $z=0$ plane. Configuration (a) has exterior 
cell-to-cell angle $\varphi = 0$  and (b) $\varphi = 36^\circ \approx 0.63 \, \text{rad}$, corresponding to a half circle and 
the maximum long-range flux. (c) Influx through a sphere of radius $R \rightarrow \infty$ (neglecting advective flux) 
shown in purple and flux through the collar of cells in green. Boundary conditions are velocity-prescribed. The graph 
ends at $\varphi = 60^\circ \approx 1.0\, \text{rad}$ corresponding to a regular hexagon.}
\label{fig:chainflow}
\end{figure*}

\section{Diffusion effects}

The flux calculations above were done in the limit of infinite P{\'e}clet number, \textit{i.e.} ignoring diffusion.
This includes ignoring the effect of crowding:
one cell eating leaves less food in the area for its neighbors.
To study these effects we consider the axially symmetric pole-to-pole arrangement shown in Fig. \ref{fig:dimerflow}b, 
the system which has the highest stresslet flux, and compare it to the single-celled swimmer (which is also axially symmetric).
The nutrient field $c(\x)$ obeys the advection-diffusion equation
\begin{align}
D \, \nabla^2 c - \u \cdot \nabla c= -R(\x)
\end{align}
with sinks $R(\x)$ at the position of the collars: $R(\x) = \sum_k R_k \, \delta(\x-\x_k)$.
By posing the problem in a weak formulation with no-flux conditions at the sphere boundaries, we obtain
\begin{align} \nonumber
\int_\Omega \Big[ & D \rho \frac{\partial c}{\partial \rho} \frac{\partial q}{\partial \rho}  + D \rho \frac{\partial c}{\partial x} \frac{\partial q}{\partial x} + \rho \, u_\rho \frac{\partial c}{\partial \rho} q + \rho \, u_x  \frac{\partial c}{\partial x} q \Big] \, \d \x \\
&= \sum_k Z_k \, q(\x_k) \quad \forall q \in Q,
\label{eq:fem}
\end{align} 
where $q$ is a test function from some Sobolev space $Q$, we have replaced $y$ by $\rho$ to make explicit the use of cylindrical coordinates, and $Z_k = \rho R_k$ such that $Z = \sum_k Z_k$ is representative for the nutrient uptake of the axially symmetric sinks.

Far away from the colony we require the nutrient field to be 
unaffected by the colony, and thus have the boundary condition $c(r) \rightarrow c_0$ as $r \rightarrow \infty$. 
Diffusion-dominated decay to $c_0$ will be of the form $c - c_0 \sim r^{-1}$, 
but for swimming colonies and large P{\'e}clet numbers, advection will dominate even far from the colony.
Using a custom mesher, we thus triangulate a massive domain ($\sim 10$ mm) with increasing detail close to the colony and take $c = c_0$ at this boundary.
Taking the choanoflagellates to be perfect eaters, the values of $Z_k$ can be calculated by imposing $c(\x)=0$ on the collars.
We solve Eq. \eqref{eq:fem} by implementing the finite element method. 
The velocity field is taken from the boundary element simulation and projected onto a solenoidal field to prevent finite numerical compressibility.

\begin{figure*}
\centering
\includegraphics{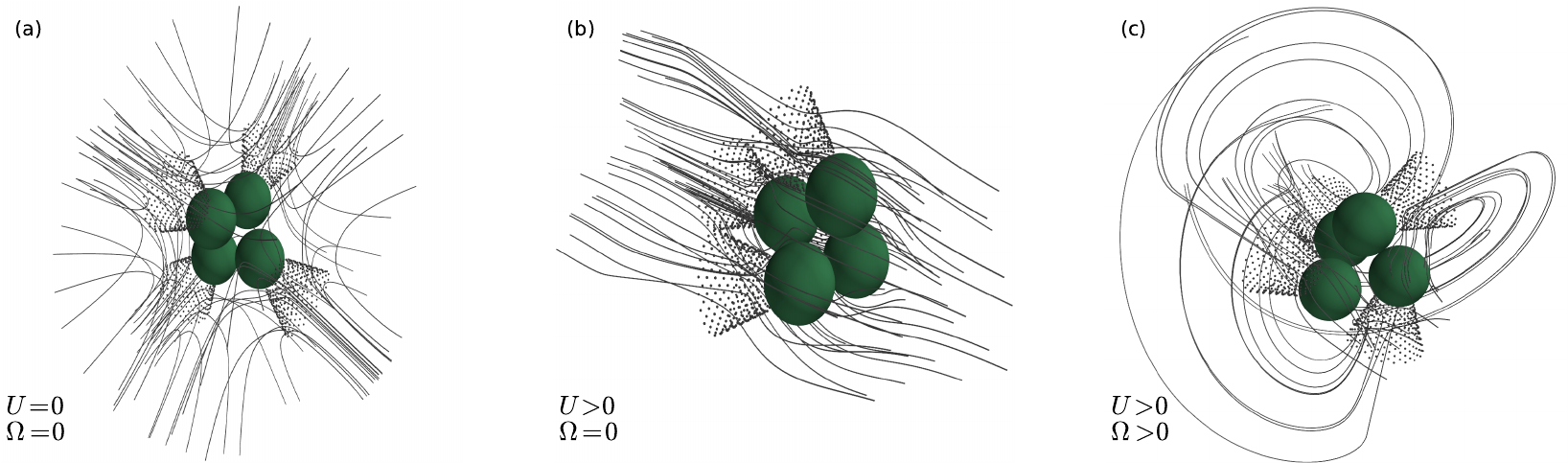}
\caption{Tetrahedron rosette colonies and streamlines of the surrounding flow. (a) All flagella pointing outwards. (b) All flagella pointing approximately in the same direction, making the colony swim faster than a unicell. (c) One flagellum propelling the colony, the remaining rotating it.}
\label{fig:tetra}
\end{figure*}

\begin{figure}[b]
\centering
\includegraphics{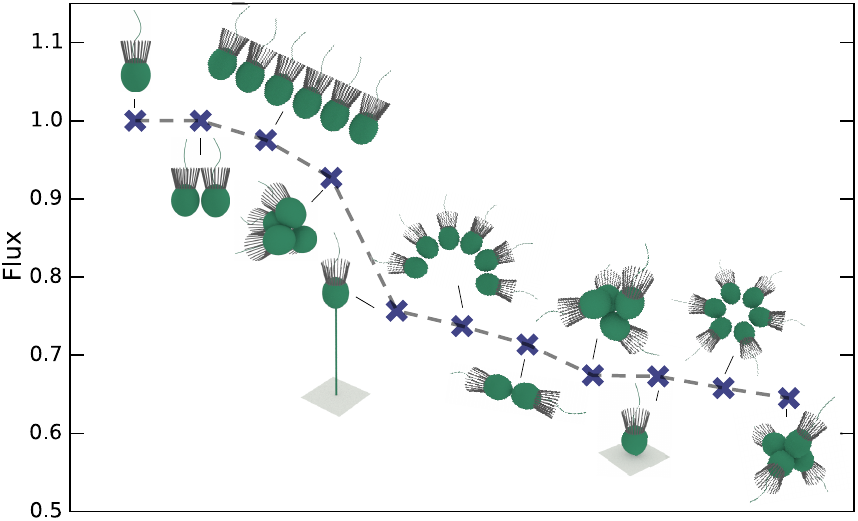}
\caption{Overview of flux across collars on the various considered configurations in no external flow with velocity-prescribed boundary conditions.
All normalized by the flux of the swimming unicell. 
From left to right: swimming unicell, side-by-side dimer, straight chain, straight rosette, long-stalk thecate cell, semi-circle chain,
pole-to-pole dimer, rotating rosette, short-stalk thecate cell, hexagonal chain, parallel rosette.}
\label{fig:fluxconfigs}
\end{figure}

In general, including diffusion increases the nutrient uptake for both the swimming unicell and the dimer colony.
The quantity of interest is the relative nutrient uptake of the cells in the colony to that of the single swimming cell.
Fig. \ref{fig:nutrientfield} shows the solution $c(\x)$  with $c_0 = 1$ for (a) the swimming unicell and (b) the pole-to-pole colony at two different values of diffusion constant $D$.
The diffusion coefficient of a passive nutrient, such as non-motile bacteria, can be calculated from the Stokes-Einstein relation $D = k_B T / 6 \pi \mu a$, where $a$ is the effective radius of the nutrient.
For typical prey such as \textit{Algoriphagus machopongenesis} this yields $D = 0.5 \, \upmu \text{m}^2/$s \cite{Roper2013}, and increases for smaller prey.
For motile prey the diffusion constant can be much larger, since it is enhanced by swimming.
For an organism swimming with speed $v$ and with rotational diffusion constant $D_r$, the effective diffusion constant 
scales as $D \sim v^2/D_r$.
Thus for $v \sim 10 \, \upmu$m/s and $D_r \sim 0.1 \, \text{s}^{-1}$, $D \sim 10^2 - 10^3 \, \upmu \text{m}^2/s$.
Moreover, even for non-motile prey, the surrounding fluid environment may be inhomogeneous and noisy, 
and such noisy flow can heuristically be associated with an increased diffusion constant.
Overall the prey diffusion constant can vary over several orders of magnitude.

Figure \ref{fig:nutrientfield}c shows how increasing the diffusion constant from zero gives a decrease in feeding 
of the colony compared to the swimming unicell at small diffusion constants.
This is due to the fact that as the effects of diffusion is increased, the fluid flux across the collar of the non-swimming colony is no longer pristine; that is the nutrients/prey of the fluid crossing the collar in steady state have already been partly consumed.
Swimming counteracts this effect, and accordingly the unicell is not affected significantly by this.
As the diffusion constant becomes large, the effects of advection diminish.
This regime is shown in the inset of Fig. \ref{fig:nutrientfield}c.
In the limit $D \rightarrow \infty$ (dashed line), the effects of advection can be ignored,
and with it the difference due to flow produed by the unicell and the colony.
However, also in this limit the unicell outperforms the colony, since there is a reduction in feeding  due to the sharing of prey between cells in a colony.

The importance of advective fluxes due to swimming depends on the opening angle of the collar.
The solid curve of Fig. \ref{fig:nutrientfield}c corresponds to the angle, $20^\circ$, shown in (a,b), the dashed curves show the result for smaller opening angles.
The top (orange) curve is for straight collars, and even in this case is it quite advantageous to be swimming.

\section{Larger Colonies}

Colonies of \textit{S. rosetta} exist with both chain and rosette morphologies.
From the above study of dimers, we expect the collar fluid flux to be maximized for a straight chain of cells.
Fig. \ref{fig:chainflow} shows the result on chains of six cells with varying exterior cell-to-cell angle $\varphi$ --- from straight to regular hexagonal shapes.
Fig. \ref{fig:chainflow}a shows the flow around the straight configuration ($\varphi=0$), and Fig. \ref{fig:chainflow}b shows a semicircle ($\varphi \approx 0.63$).
For the long-range flux, shown by the purple curve in Fig. \ref{fig:chainflow}c, we find as in Ref. \cite{Roper2013} that the semicircle configuration is the global maximum.
This rich behavior of the long-range flux disappears completely in the collar flux as shown by the green curve in Fig. \ref{fig:chainflow}c,
and again we find that the globally optimal configuration is the one that swims the fastest: the straight chain.

It was suggested \cite{Roper2013} that while the long-range flux increase appears for chain morphologies, rosette-shaped colonies will not have this effect.
To exemplify rosette colonies, we take a tetrahedron of cells and consider three distinct flagella configurations, 
the resulting flow fields of which are shown in Fig. \ref{fig:tetra}:
(a) flagella pointing outwards parallel with the line from the center-of-mass to the cell,
(b) flagella pointing approximately in the same direction, and
(c) one flagellum propelling the colony and the remaining three rotating it.
Configuration (a) will not swim nor rotate. The flagella of configuration (b) point almost in the same direction, except for a small angle to make sure collars do not overlap.
This configuration swims faster than a unicell due to reduced drag from the tetrahedron configuration.
The collars of a rotating colony will sweep a larger volume, which is exemplified by configuration (c).
In terms of the flux across the collar, we compare in Fig. \ref{fig:fluxconfigs} these tetrahedra to the other 
morphologies considered.
The non-swimming tetrahedron is the worst of all considered configurations.
With all considered configurations we have found the fastest swimmer to also have the highest collar flux.
This does not hold for configuration (b), however. Although it swims $\sim 20\, \%$ faster than a unicell, the middle collar is confined between the other collars and accordingly has a significantly reduced flux.
The rotating colony (c) is also in the lower end of the collar flux. Since the flagella are already drawing the surrounding fluid through the collars, 
the extra volume swept by rotating makes no difference --- one side of the collar will have an increased flux, but the opposing side will be equally reduced.

\section{Thecate cells}
\begin{figure}[tb]
\centering
\includegraphics{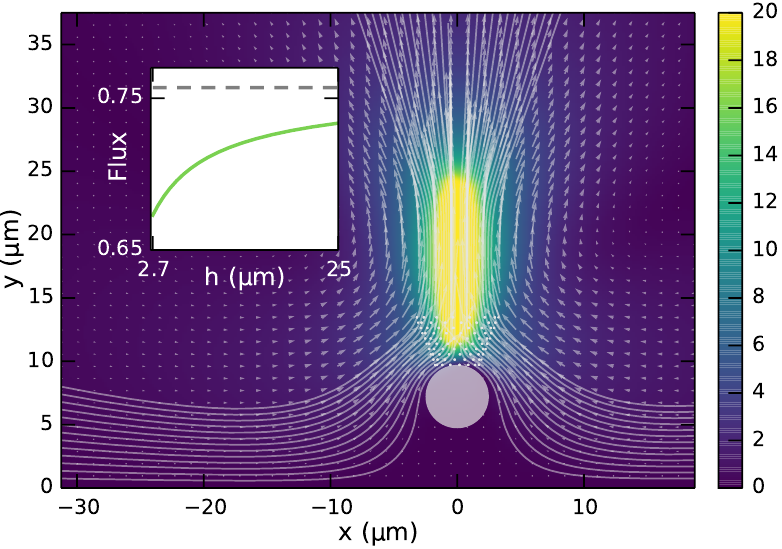}
\caption{Thecate cell above wall at $y=0$. Background color and vector field shows the velocity field, 
with color scale in units of $\upmu$m/s. Streamlines are at $z=0$.
Inset shows collar flux normalised by that of a swimming unicell as a function of height $h$ above the wall.}
\label{fig:thecate}
\end{figure}

For completeness, we must include in this study the sessile form of \textit{S. rosetta}.
These attach to a wall by building a so-called theca.
Such single-celled sessile feeders have previously been studied \cite{Physics1979, Pepper2013}.
To account for the no-slip condition on the nearby wall, 
we add image singularity solutions to Eq. \eqref{eq:regstokeslet} at the mirror point over the wall.
For a singular Stokeslet, the images that give no-slip on the wall are a Stokeslet of opposite sign, a potential dipole, and a Stokeslet doublet \cite{Blake1971, Pozrikidis1992}.
Similar to the unbounded version, Eq. \eqref{eq:regstokeslet}, a regularized version is known \cite{Ainley2008}, which we exploit and replace $\G$ (in Appendix Eq. \eqref{eq:vbem2}) with a tensor including these images.
Fig. \ref{fig:thecate} shows the resulting flow.
We consider straight thecate cells, which is the configuration with highest flux, 
although in the absence of external flow, feeding at an angle can be advantageous 
in order to reduce recirculating eddies \cite{Pepper2013}.
In the inset of Fig. \ref{fig:thecate} the collar flux is plotted as a function of the height $h$ above the wall that the cell is attached to.
Overall the flux is reduced compared to the swimming unicell, but this results only holds in the absence of external flow.
Being stuck to the wall, thecate cells gain an advantage from external flows that suspended cells do not.
As long as the external flow is comparable to or larger than the swimming speed of a unicell, the thecate form becomes advantageous.
Not surprisingly, Fig. \ref{fig:thecate} shows that placing the cell further away from the wall increases the flux; 
this is the very reason that the cells build a stalk on the theca.
The difference in terms of flux between no stalk and an infinitely long stalk (dashed line in Fig. \ref{fig:thecate} inset) is about $10 \, \%$ of the flux of the swimming unicell.

\section{Conclusions}

We have found that swimming is the best strategy to maximize the prey flux across the feeding collar in
choanoflagellates, in agreement with the result found for absorbing feeders modelled as squirmers \cite{Lauga2011a}, 
and that there is no hydrodynamic feeding advantage for colonies compared to single cells.
With flagella orientations parallel to cell positions in rosette colonies, the swimming speeds will be significantly lowered.
However, real rosette colonies tend to swim at speeds that are comparable to unicell slow-swimmers \cite{Dayel2011, Kirkegaard2016}.
One might hypothesize that the advective flux is a selection factor for flagellar orientations that allow swimming.
Swimming moreover is a natural method for replenishing the surrounding fluid, and as discussed, thereby limit the hindering effects of diffusion.
Prey trajectories are more aligned with the collar for swimming cells than for stationary cells, and
if the capture probability decreases with alignment (\textit{e.g.} if prey bounce off the collar) this will favor stationary cells.
But live imaging is needed to asses the magnitude of such an effect, and it would have to be very large 
in order to give to colonies the overall advantage.
For \textit{S. rosetta} the fact that colonies tend to form when a culture is kept in log phase,
\textit{i.e.} with plenty of nutrients, suggests that enhanced feeding efficiency per se 
is not a driving force behind colony formation,
and other factors such as size as a prevention against predation could be more important.  Taken together 
with the fact that a molecular
species released by certain prey bacteria triggers the formation of the multicellular form \cite{Alegado2012} suggests
that the driving forces behind transitions to multicellularity are subtle indeed.

\begin{acknowledgments}We thank Fran{\c c}ois J. Peaudecerf and Pierre A. Haas for discussions.
This work was supported in part by the EPSRC and St. Johns College (JBK), and a Wellcome Trust Senior Investigator Award 
(REG).
\end{acknowledgments}

\section{Appendix: numerical method}
The Stokes flow around thin elements such as straight lines is often described by slender body theory. 
Utilizing the present framework they may also be described as in Eq. \eqref{eq:regstokeslet} by a line 
integral of regularized Stokeslets with $\epsilon$ suitably chosen to model the thickness of the line \cite{Smith2009}.
Instead of discretizing $\f(\x)$ over triangular elements, for example, we parametrize $\f(\x)$ on the spheres in terms of spherical harmonics and on the lines with Legendre polynomials.
On a sphere $S_i$ we thus write
\begin{equation}
f^{S_i}_j = \sum_{l=0}^{\infty} \sum_{m=-l}^{l} c^{ij}_{lm} Y_{lm}(\phi,\theta),
\end{equation}
where $Y_{lm}$ are the real spherical harmonics defined in terms of the conventional spherical harmonics as
\begin{equation}
Y_{lm} = \begin{cases} 
	  Y^l_m & m=0 \\
      (-1)^m \sqrt{2} \, \text{Im}[Y^l_m] &  m<0 \\
      (-1)^m \sqrt{2} \, \text{Re}[Y^l_m] &  m>0.
   \end{cases}
\end{equation}
And on line $\ell_i$ we write
\begin{equation}
f^{\ell_i}_j = \sum_{n=0}^{\infty} c^{ij}_{n} P_{n}(s),
\end{equation}
where $P_n$ is the $n$-th Legendre polynomial.
Eq. \eqref{eq:bem1} thus becomes
\begin{align} \nonumber
\u(\x)_j &= \sum_{i=1}^{n_S}  \sum_{l=0}^{\infty} \sum_{m=-l}^{l}  c^{ik}_{lm}  \int_{-\pi}^{\pi} \d \phi  \int_0^\pi \sin \theta \, \d \theta   \\ \nonumber
& \quad \quad G_{jk}(\x-[\r_i + a \y(\phi,\theta)]) Y_{lm}(\phi,\theta)   \\
& +\sum_{i=1}^{n_\ell}  \sum_{n=0}^{\infty} c^{ik}_{n}  \int_{-1}^1 \d s \,  G_{jk}(\x-\y_i(s)) P_{n}(s)  ,
\label{eq:vbem2}
\end{align}
where the Einstein summation is implied over $k$ and
\begin{equation}
\y(\phi,\theta) = \begin{pmatrix}
\sin \theta \cos \phi\\
\sin \theta \sin \phi\\
\cos \theta
\end{pmatrix},
\label{eq:y}
\end{equation}
spans a sphere such that $\r_i + a \y(\phi,\theta)$ is a sphere of radius $a$ centred on $\r_i$. The flagella lines are spanned by 
\begin{equation}
\y_i(s) = {\boldsymbol \ell}_i + \frac{s+1}{2} \, \, {\boldsymbol d}_i, \, s\in [-1,1],
\end{equation}
where ${\boldsymbol \ell}_i $ is the base position, $\hat{{\boldsymbol d}_i}$ its orientation, and $|{\boldsymbol d}_i|$ its length.
Truncating the spherical harmonic expansion at $l = n_Y$ and the Legendre expansion at $n=n_P$, we have $3 n_S (1+n_H)^2 + 3 n_\ell (n_P+1)$ unknown coefficients to determine. 
The integrals must be evaluated numerically. Gauss-Legendre quadrature enables exact numerical integration of polynomials, and for other functions gives good approximations to the integrals by
\begin{equation}
\int f(s) \, \d s \simeq \sum_i w^\ell_i f(s_i),
\end{equation}
where $w^\ell_i$ are weights associated with the quadrature points $s_i$.
Likewise, Lebedev quadrature enables exact numerical integration of spherical harmonics. Thus spherical integrals can be numerically approximated as
\begin{equation}
\iint f(\phi,\theta) \sin \theta \, \d \theta \, \d \phi \simeq \sum_i w^S_i f(\phi_i, \theta_i),
\end{equation}
where $w^S_i$ are weights associated with the quadrature points $(\theta_i, \phi_i)$. The numerical schemes become exact if $f$ can be expanded precisely up to some order by using an appropriate number of quadrature points. 

Neutrally buoyant microorganisms, of which choanoflagellates are good approximations, are furthermore 
force- and torque-free. Therefore,
\begin{align} \label{eq:noforce}
&\sum_{i=1}^{n_S} \int_{S_i} \f^{S_i}(\x) \, \d S  + \sum_{i=1}^{n_\ell} \int_{\ell_i} \f^{\ell_i}(\x) \, \d 
\ell = 0, \\ \label{eq:notorque}
&\sum_{i=1}^{n_S} \int_{S_i} \x \times \f^{S_i}(\x) \, \d S  + \sum_{i=1}^{n_\ell} \int_{\ell_i} \x \times \f^{\ell_i}(\x) \, \d 
\ell = 0.
\end{align}
These equations set the swimming velocity $\U$ and rotational velocity $\O$ such that swimming drag forces and torques precisely cancel the propulsive ones.
In the frame of reference of the swimming organism we thus add to Eq. \eqref{eq:vbem2} the term $(\U + \x \times \O)_j = \U_j + \epsilon_{jpq} \, \x_p \, \O_q$, where $\epsilon_{jpq}$ is the Levi-Civita symbol. In terms of the coefficients $\{ c \}$, Eq. \eqref{eq:noforce} becomes
\begin{align}
& 2\sqrt{\pi} \, \sum_{i=1}^{n_S}  c^{ij}_{00} + 2 \sum_{i=1}^{n_\ell} c^{ij}_0 = 0
\label{eq:coeffnoforce}
\end{align}
and Eq. \eqref{eq:notorque}
\begin{align} \nonumber
& \sum_{i=1}^{n_S}  \epsilon_{jpq} \left[ 2 \sqrt{\pi}  \, (\r_i)_p \, c^{iq}_{00} + 2 a \, \sqrt{\frac{\pi}{3}} \, c^{iq}_{1,m(q)} \right] \\
& +  \, \sum_{i=1}^{n_\ell} \epsilon_{jpq} \left[ [2 ({\bs \ell}_i)_p + ({\bs d}_i)_p] \, c_0^{iq} + \frac{1}{3} ({\bs d}_i)_p \, c_1^{iq} \right] = 0,
\label{eq:coeffnotorque}
\end{align}
where $m(1) = 1, m(2)=-1, m(3)=0$.

By choosing the same of number collocation points $\{ \x_i \}$ for evaluating the velocities $\{ \u(\x_i) \}$ 
as the total number of coefficients $\{ c \}$ the linear system of equations can be solved for $\{ c \}$, $\U$ and $\O$.
By exploiting orthogonality, we could expand $\u$ on the spheres and lines in terms of spherical harmonics and Legendre polynomials, respectively. 
However, for the systems considered here the computational bottle neck is the Gaussian quadratures, the number of which would be squared if this method were employed.
\vspace{2.0em}

\end{document}